\documentclass[aps,twocolumn,pre,amsmath]{revtex4}
\usepackage{graphicx}

\begin{document}

\title{Maier-Saupe-type theory of ferroelectric nanoparticles in nematic
liquid crystals}

\author{Lena M. Lopatina}
\affiliation{Liquid Crystal Institute, Kent State University, Kent, OH 44242,
USA}
\author{Jonathan V. Selinger}
\email{jselinge@kent.edu}
\affiliation{Liquid Crystal Institute, Kent State University, Kent, OH 44242,
USA}

\date{May 17, 2011}

\begin{abstract}

Several experiments have reported that ferroelectric nanoparticles have
drastic effects on nematic liquid crystals---increasing the isotropic-nematic
transition temperature by about 5 K, and greatly increasing the sensitivity to
applied electric fields.  In a recent paper [L. M. Lopatina and J. V.
Selinger, Phys.\ Rev.\ Lett.\ \textbf{102}, 197802 (2009)], we modeled these
effects through a Landau theory, based on coupled orientational order
parameters for the liquid crystal and the nanoparticles.  This model has one
important limitation:  Like all Landau theories, it involves an expansion of
the free energy in powers of the order parameters, and hence it overestimates
the order parameters that occur in the low-temperature phase.  For that
reason, we now develop a new Maier-Saupe-type model, which explicitly shows
the low-temperature saturation of the order parameters.  This model reduces to
the Landau theory in the limit of high temperature or weak coupling, but shows
different behavior in the opposite limit.  We compare these calculations with
experimental results on ferroelectric nanoparticles in liquid crystals.

\end{abstract}
\maketitle

\section{Introduction}

One important goal of modern liquid-crystal research is to enhance the
properties of liquid crystals through \emph{physical} methods, without the
need for new chemical synthesis.  One way to achieve this goal is to put
colloidal particles into liquid crystals.  If the particles have a length
scale of microns, they induce elastic distortions of the liquid crystals, and
these distortions mediate an effective interaction between the particles.  The
particles may then form a periodic array, leading to a composite material with
potential applications in
photonics~\cite{poulin97,gu00,stark01,yada04,smalyukh05,musevic06}.  By
comparison, if the particles have a length scale of 10--100 nm, they are too
small to distort the liquid crystal.  In that case, the system enters into
another regime of behavior, in which the particles function as molecular
additives to change the effective properties of the liquid-crystal host.  One
particularly interesting case occurs if the particles are
\emph{ferroelectric}.  Experiments have shown that low concentrations of
ferroelectric Sn$_2$P$_2$S$_6$ or BaTiO$_3$ nanoparticles increase the
orientational order parameter, increase the isotropic-nematic transition
temperature, and decrease the switching voltage for the Frederiks
transition~\cite{reznikov03,ouskova03,reshetnyak04,buchnev04,glushchenko06,%
reshetnyak06,li06,buchnev07}.  Thus, they provide a new opportunity to enhance
the properties of liquid crystals for technological applications.

To make further progress with these materials, it is essential to develop a
theory for the interaction between liquid crystals and ferroelectric
nanoparticles.  In previous theoretical research, Reshetnyak \emph{et al.}
have developed a theoretical approach based on
electrostatics~\cite{reshetnyak06,li06,shelestiuk11}.  In this theory, the key
issue is how an ensemble of nanoparticles with aligned dipole moments can
polarize the liquid-crystal molecules, hence increasing the intermolecular
interaction.  This electrostatic effect enhances the isotropic-nematic
transition temperature and reduces the Frederiks transition voltage.  In
related research, Pereira \emph{et al.} have performed molecular dynamics
simulations of ferroelectric nanoparticles immersed in a nematic liquid
crystal~\cite{pereira10}.  These simulations also assume that the
nanoparticles are aligned, and they also find a substantial enhancement of
liquid-crystal order.

In a recent paper~\cite{lopatina09}, we proposed a different type of theory
for the statistical mechanics of ferroelectric nanoparticles in liquid
crystals.  In that theory, we suppose that \emph{both} the liquid crystals and
the nanoparticles have distributions of orientations, as illustrated in
Fig.~1.  These distributions are characterized by two orientational order
parameters, which interact with each other.  Using a Landau theory, we showed
that the coupling stabilizes the nematic phase.  By estimating the strength of
the coupling, we calculated the enhancement in the isotropic-nematic
transition temperature.  We also predicted that the nanoparticles would
greatly increase the Kerr effect, the response of the isotropic phase to an
applied electric field.

\begin{figure}[b]
\includegraphics[width=2.25in]{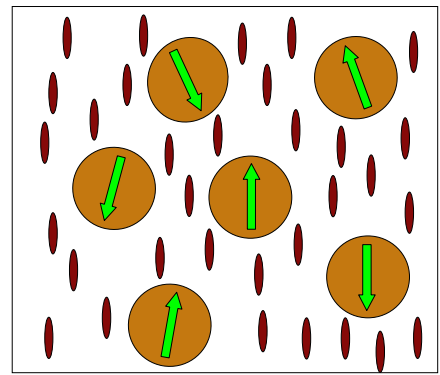}
\caption{(Color online) Schematic illustration of ferroelectric nanoparticles
suspended in a liquid crystal.  The electrostatic dipole moments of the
nanoparticles have a distribution of orientations.}
\end{figure}

Although the work of Ref.~\cite{lopatina09} demonstrates an important physical
mechanism, we must acknowledge that it has one mathematical limitation:  Like
all Landau theories, it involves an expansion of the free energy in powers of
the order parameters.  This expansion is valid when the order parameters are
small, but it breaks down when they become large.  In particular, the theory
allows the order parameters to become larger than 1, which is clearly
impossible.  For ferroelectric nanoparticles in a liquid crystal, the
nanoparticle order parameter is not necessarily small, even near the
isotropic-nematic transition.

The purpose of the current paper is to generalize the previous theory by
eliminating the assumption that the order parameters are small.  For this
generalization, we now use a Maier-Saupe-type theory instead of a Landau
theory.  We still consider the same physical concept of coupled orientational
order parameters for the liquid crystals and the nanoparticles, and we still
use the same energy of interaction between them.  However, we now use a more
general expression for the entropy, not a power series, which enforces the
constraint that the order parameters cannot become larger than 1.  This change
allows us to avoid the potential mathematical inconsistency of Landau theory.

Like our previous calculation, the work presented here shows that doping
liquid crystals with ferroelectric nanoparticles enhances the
isotropic-nematic transition temperature.  In the limit of weak coupling
between the nanoparticles and the liquid crystal, the Maier-Saupe-type theory
exactly reduces to the Landau theory.  However, in the case of strong
coupling, the new theory predicts a smaller but still substantial enhancement.
Rough estimates suggest that the experimental system is in the limit of strong
coupling, so it is important to use this modified theory.  Furthermore, the
work presented here also predicts the Kerr effect as a function of applied
electric field.  In the limit of low electric field, the Maier-Saupe-type
theory exactly reduces to the Landau theory.  However, for larger field, the
nanoparticle order saturates and the enhanced Kerr effect is cut off.

The plan of this paper is as follows.  In Sec.~II we present the formalism of
Maier-Saupe theory, with interacting orientational distributions for
liquid-crystal molecules and nanoparticles.  In Sec.~III we apply this
formalism to calculate the isotropic-nematic transition temperature, and
determine the enhancement due to nanoparticles.  In Sec.~IV we use the same
formalism to calculate the Kerr effect of induced orientational order under an
applied electric field, and investigate how this effect depends on the
magnitude of the field.  Finally, in Sec.~V we discuss the main conclusions of
this study.

\section{Overview of Maier-Saupe Theory}

In this section we introduce the free energy for a system of liquid-crystal
molecules with ferroelectric nanoparticles.  To construct the free energy, we
use the fundamental equation of mean-field theory,
\begin{equation}
F=\langle H\rangle+k_\mathrm{B}T\langle\ln\varrho\rangle,
\end{equation}
where the first term is the energy, the second term is the entropic
contribution to the free energy, and the averages are taken over the
single-particle distribution function $\varrho$.  Hence, the first step is to
define the distribution functions for liquid-crystal molecules and
nanoparticles.

Liquid-crystal molecules are rod-shaped objects, with each molecule
characterized by the direction of its long axis $\mathbf{m}$.  In the nematic
phase, these axes are preferentially oriented along the average director
$\mathbf{n}$.  Because the individual molecules are equally likely to point
along $+\mathbf{n}$ or $-\mathbf{n}$, the single-molecule distribution
function can be written as
\begin{equation}
\varrho_\mathrm{LC}(\theta)=\frac{\exp({U_\mathrm{LC}P_2(\cos\theta))}}
{\int_{-1}^{1}d(\cos\theta)\exp({U_\mathrm{LC}P_2(\cos\theta))}},
\label{LCdistribution}
\end{equation}
where $\theta$ is the angle between the molecular orientation $\mathbf{m}$ and
the average director $\mathbf{n}$, and $P_2$ is the second Legendre
polynomial.  The parameter $U_\mathrm{LC}$ is a variational parameter, which
acts as an effective field on the molecular orientation.  It is related to the
standard nematic order parameter
$S_\mathrm{LC}=\langle P_2(\cos\theta)\rangle$ by
\begin{equation}
S_\mathrm{LC}=\frac{\int_{-1}^{1}d(\cos\theta)P_2(\cos\theta)
\exp({U_\mathrm{LC}P_2(\cos\theta))}}
{\int_{-1}^{1}d(\cos\theta)\exp({U_\mathrm{LC}P_2(\cos\theta))}}
\label{S_LC}
\end{equation}
Note that $U_\mathrm{LC}$ ranges from 0 to $\infty$, while $S_\mathrm{LC}$
ranges from 0 to 1.

We can now calculate the free energy of a pure liquid-crystal system.
Maier-Saupe theory assumes that the interaction energy between neighboring
molecules $i$ and $j$ is proportional to $-(\mathbf{m}_i\cdot\mathbf{m}_j)^2$.
With the assumed distribution function, the average interaction energy becomes
\begin{equation}
F_\mathrm{energetic}^\mathrm{LC}=\langle H\rangle=-\frac{1}{3}J N_\mathrm{LC}
S_\mathrm{LC}^2,
\end{equation}
where $N_\mathrm{LC}$ is the number of liquid-crystal molecules in the system,
and $J$ is an energetic parameter proportional to the interaction strength and
the number of neighbors per molecule.  Furthermore, the entropic contribution
to the free energy can be written as
\begin{eqnarray}
F_\mathrm{entropic}^\mathrm{LC}&=&k_\mathrm{B}T N_\mathrm{LC}
\langle\ln\varrho_\mathrm{LC}\rangle\nonumber\\
&=&k_\mathrm{B}T N_\mathrm{LC}\Bigl[U_\mathrm{LC}S_\mathrm{LC}\\
&&-\ln\left[\textstyle\int_{-1}^{1}d(\cos\theta)
\exp(U_\mathrm{LC}P_2(\cos\theta))\right]\Bigr].\nonumber
\end{eqnarray}
By combining these pieces, we obtain the total free energy of the liquid
crystal,
\begin{eqnarray}
\label{purefreeenergy}
\frac{F}{N_\mathrm{LC}k_\mathrm{B}T}&=&
-\frac{J}{3k_\mathrm{B}T}S_\mathrm{LC}^2+U_\mathrm{LC}S_\mathrm{LC}\\
&&-\ln\left[\textstyle\int_{-1}^{1}d(\cos\theta)
\exp(U_\mathrm{LC}P_2(\cos\theta))\right]\nonumber
\end{eqnarray}

The free energy of Eq.~(\ref{purefreeenergy}) is a function of the temperature
$T$ and the variational parameter $U_\mathrm{LC}$, with $S_\mathrm{LC}$
defined implicitly as a function of $U_\mathrm{LC}$ through Eq.~(\ref{S_LC}).
By minimizing the free energy over $U_\mathrm{LC}$ for varying temperature, we
can find the liquid crystal has a first-order transition from the isotropic
phase with $U_\mathrm{LC}=S_\mathrm{LC}=0$ to the nematic phase with
\begin{equation}
U_\mathrm{LC}=1.95, \quad S_\mathrm{LC}=0.429.
\label{ULCpure}
\end{equation}
The numerical solution for the transition temperature in this pure liquid
crystal is
\begin{equation}
T_\mathrm{NI}=0.147\frac{J}{k_\mathrm{B}}.
\label{TNIJ}
\end{equation}
Also, we can find an analytic solution for the limit of supercooling,
\begin{equation}
T^* =\frac{2J}{15k_\mathrm{B}}=0.133\frac{J}{k_\mathrm{B}}.
\label{tstar}
\end{equation}
From experiments we know $T^*$ and $T_\mathrm{NI}$ for any particular
liquid-crystal material, so we can use Eq.~(\ref{TNIJ}) or~(\ref{tstar}) to
determine $J$ for that material,
\begin{equation}
J=6.81 k_\mathrm{B} T_\mathrm{NI}.
\end{equation}

Once we add nanoparticles to the system, we get another distribution function
for the orientations of the nanoparticle dipole moments.  By symmetry, we
expect that this distribution should be aligned along the same axis
$\mathbf{n}$ as the liquid-crystal distribution.  However, the magnitude of
the order may be different.  Hence, we can write the nanoparticle distribution
function as
\begin{equation}
\varrho_\mathrm{NP}(\theta)=\frac{\exp({U_\mathrm{NP}P_2(\cos\theta))}}
{\int_{-1}^{1}d(\cos\theta)\exp({U_\mathrm{NP}P_2(\cos\theta))}},
\label{NPdistribution}
\end{equation}
where $U_\mathrm{NP}$ is a variational parameter for the nanoparticles.  The
orientational order parameter $S_\mathrm{NP}$ of the nanoparticles can be
defined by analogy with the liquid-crystal order parameter as
\begin{equation}
S_\mathrm{NP}=\frac{\int_{-1}^{1}d(\cos\theta)P_2(\cos\theta)
\exp({U_\mathrm{NP}P_2(\cos\theta))}}
{\int_{-1}^{1}d(\cos\theta)\exp({U_\mathrm{NP}P_2(\cos\theta))}}.
\label{S_NP}
\end{equation}
Just as in the liquid-crystal case, note that $U_\mathrm{NP}$ ranges from 0 to
$\infty$, while $S_\mathrm{NP}$ ranges from 0 to 1.

As we discussed in our previous paper~\cite{lopatina09}, the ferroelectric
nanoparticles create static electric fields, which interact with the
dielectric anisotropy of the liquid crystal.  By averaging the interaction
energy over the distribution functions $\varrho_\mathrm{LC}$ and
$\varrho_\mathrm{NP}$, we obtain
\begin{equation}
F_\mathrm{interaction}=-K_\mathrm{NP}N_\mathrm{NP}S_\mathrm{LC}S_\mathrm{NP}.
\label{NPenergy}
\end{equation}
In this expression, $N_\mathrm{NP}$ is the number of nanoparticles in the
system, and $K_\mathrm{NP}$ is an energetic parameter representing the
strength of the interaction.  For an unscreened electrostatic interaction, we
derived
\begin{equation}
K_\mathrm{NP}=\frac{\varepsilon_0 \Delta\varepsilon p^2}
{180\pi(\varepsilon_0\varepsilon)^2 R^3}
=\frac{4\pi\varepsilon_0 \Delta\varepsilon P^2 R^3}
{405(\varepsilon_0 \varepsilon)^2}.
\end{equation}
where $p$, $P$, and $R$ are dipole moment, polarization, and radius of a
nanoparticle, and $\varepsilon$ and $\Delta\varepsilon$ are the dielectric
constant and dielectric anisotropy of the bulk liquid crystal.  If the
interaction is screened by counterions, then $K_\mathrm{NP}$ is somewhat
reduced, but it is still substantial as long as the Debye screening length is
greater than the nanoparticle radius.  Hence, orientational order of the
liquid-crystal molecules tends to favor orientational order of the
nanoparticles, and vice versa.

Whenever there is an aligning effect, there must be an entropic cost.  By
analogy with the entropic term for liquid-crystal molecules, the entropic
penalty for aligning the nanoparticles is
\begin{eqnarray}
F_\mathrm{entropic}^{NP}&=&k_\mathrm{B}T N_\mathrm{NP}
\langle\ln\varrho_\mathrm{NP}\rangle\nonumber\\
\label{NPentropy}
&=&k_\mathrm{B}T N_\mathrm{NP}\Bigl[U_\mathrm{NP}S_\mathrm{NP}\\
&&-\ln\left[\textstyle\int_{-1}^{1}d(\cos\theta)
\exp(U_\mathrm{NP}P_2(\cos\theta))\right]\Bigr].\nonumber
\end{eqnarray}

The total free energy for liquid-crystal molecules and nanoparticles is now
the combination of Eqs.~(\ref{purefreeenergy}), (\ref{NPenergy}),
and~(\ref{NPentropy}),
\begin{eqnarray}
\frac{F}{N_\mathrm{LC}k_\mathrm{B}T}&=&
-\frac{J}{3 k_\mathrm{B}T} S_\mathrm{LC}^2
-\frac{\nu K_\mathrm{NP}}{k_\mathrm{B} T}S_\mathrm{LC}S_\mathrm{NP}\nonumber\\
\label{freeenergy}
&&+U_\mathrm{LC}S_\mathrm{LC}+\nu U_\mathrm{NP}S_\mathrm{NP}\\
&&-\ln\left[\textstyle\int_{-1}^{1}d(\cos\theta)
\exp(U_\mathrm{LC}P_2(\cos\theta))\right]\nonumber\\
&&-\nu\ln\left[\textstyle\int_{-1}^{1}d(\cos\theta)
\exp(U_\mathrm{NP}P_2(\cos\theta))\right].\nonumber
\end{eqnarray}
Note that we have normalized this free energy by the number of liquid-crystal
molecules, not by the number of nanoparticles.  For that reason, all the
nanoparticle terms in Eq.~(\ref{freeenergy}) contain a factor of
$\nu=N_\mathrm{NP}/N_\mathrm{LC}$, the ratio of the number of nanoparticles to
the total number of liquid-crystal molecules.

To summarize, we have derived the free energy for the system of ferroelectric
nanoparticles suspended in a liquid crystal.  The first term represents the
aligning energy favoring orientational order of the liquid crystal, while the
second term describes the mutual aligning interaction between nanoparticle
order and liquid-crystal order.  The last terms are entropic terms that give
the free-energy penalty for any liquid-crystal or nanoparticle order.  The
free energy is a function of two variational parameters, $U_\mathrm{LC}$ and
$U_\mathrm{NP}$, and we formulate our problem as minimization over those
quantities.  Once we find them, we can calculate the order parameters
$S_\mathrm{LC}$ and $S_\mathrm{NP}$ using Eqs.~(\ref{S_LC}) and~(\ref{S_NP}).

\section{Transition Temperature}

Experiments show a substantial increase in the isotropic-nematic transition
temperature for liquid crystals doped with ferroelectric nanoparticles.  In
order to understand this phenomenon and predict how to enhance it further, we
investigate the isotropic-nematic transition using the free energy of
Eq.~(\ref{freeenergy}).

Two distinct limiting cases of this transition are possible.  If the
nanoparticle order is small, then all of the integrals in
Eq.~(\ref{freeenergy}) can be expanded in Taylor series for small
$U_\mathrm{LC}$ and $U_\mathrm{NP}$.  The expressions for the order parameters
$S_\mathrm{LC}$ and $S_\mathrm{NP}$ from Eqs.~(\ref{S_LC}) and
Eq.~(\ref{S_NP}) can also be expanded in power series in $U_\mathrm{LC}$ and
$U_\mathrm{NP}$.  Hence, the free energy can be expressed as a series in
$U_\mathrm{LC}$ and $U_\mathrm{NP}$, or equivalently as a series in
$S_\mathrm{LC}$ and $S_\mathrm{NP}$.  After some algebraic transformations, we
obtain
\begin{eqnarray}
\frac{F}{N_\mathrm{LC}k_\mathrm{B}T}&=&\mathrm{const}
+\left(\frac{5}{2}-\frac{J}{3k_\mathrm{B}T}\right)S_\mathrm{LC}^2
+\frac{5}{2}\nu S_\mathrm{NP}^2\nonumber\\
&&-\frac{\nu K_\mathrm{NP}}{k_\mathrm{B}T}S_\mathrm{NP}S_\mathrm{LC}+\dots.
\end{eqnarray}
This expression is exactly the Landau free energy as a series in the order
parameters, as discussed in our earlier paper~\cite{lopatina09}.  To find the
isotropic-nematic transition, we first minimize over $S_\mathrm{NP}$ to obtain
\begin{equation}
S_\mathrm{NP}=\frac{K_\mathrm{NP}}{5k_\mathrm{B}T}S_\mathrm{LC}.
\end{equation}
We then substitute this value into the free energy series to obtain
\begin{equation}
\frac{F}{N_\mathrm{LC}k_\mathrm{B}T}=\mathrm{const}
+\left(\frac{5}{2}-\frac{J}{3k_\mathrm{B}T}-\frac{\nu K_\mathrm{NP}^2}
{10(k_\mathrm{B}T)^2}\right)S_\mathrm{LC}^2+\dots.
\end{equation}
The change in the coefficient of $S_\mathrm{LC}^2$ shows that the
isotropic-nematic transition temperature is shifted upward by
\begin{subequations}
\label{DeltaTNI}
\begin{equation}
\Delta T_\mathrm{NI}=\frac{\nu K_\mathrm{NP}^2}
{25k_\mathrm{B}^2 T_\mathrm{NI}}
\end{equation}
In the notation of the previous paper, this shift can be written as
\begin{equation}
\Delta T_\mathrm{NI}=\frac{\pi\phi_\mathrm{NP}R^3}
{3T_\mathrm{NI}\rho_\mathrm{LC}}
\left(\frac{2\Delta\varepsilon P^2}
{675k_\mathrm{B}\varepsilon_0\varepsilon^2}\right)^2,
\end{equation}
\end{subequations}
where $\rho_\mathrm{LC}$ is the number of liquid-crystal molecules per unit
volume and $\phi_\mathrm{NP}=\frac{4}{3}\pi R^3 \rho_\mathrm{LC}\nu$ is the
volume fraction of nanoparticles.

Note that the power-series approximation works well as long as the energetic
parameter $K_\mathrm{NP}$ is small compared with $5k_\mathrm{B}T$.  In that
case the nanoparticle order parameter $S_\mathrm{NP}$ is small compared with
$S_\mathrm{LC}$, which is approximately 0.429 just below the isotropic-nematic
transition.  However, the approximation breaks down if $K_\mathrm{NP}$ becomes
large compared with $5k_\mathrm{B}T$, so that $S_\mathrm{NP}$ is large
compared with $S_\mathrm{LC}$.  In the latter case, the prediction for
$S_\mathrm{NP}$ would be greater than 0.429 on the nematic side of the
transition.  It might even be greater than 1, which would be unphysical.  This
unphysical prediction arises because the power-series expansion cannot take
account of the saturation of the order parameters at low temperatures.  Hence,
for large $K_\mathrm{NP}$ we must consider a different limiting case.

In the limit of large $K_\mathrm{NP}$, the nanoparticle order is large; i.e.
the variational parameter $U_\mathrm{NP}$ approaches infinity and the order
parameter $S_\mathrm{NP}$ approaches 1.  In that case we can approximate
Eq.~(\ref{S_NP}) to obtain
\begin{equation}
S_\mathrm{NP}=1-\frac{1}{U_\mathrm{NP}}.
\end{equation}
We can then put this approximation into the free energy of
Eq.~(\ref{freeenergy}), expand the nanoparticle entropic integral for large
$U_\mathrm{NP}$, and minimize the resulting free energy over $U_\mathrm{NP}$.
This calculation gives
\begin{subequations}
\label{MS_UNP}
\begin{eqnarray}
U_\mathrm{NP}&=&\frac{K_\mathrm{NP}}{k_\mathrm{B}T}S_\mathrm{LC},\\
S_\mathrm{NP}&=&1-\frac{k_\mathrm{B}T}{K_\mathrm{NP}S_\mathrm{LC}}.
\end{eqnarray}
\end{subequations}
Note that this calculation is self-consistent, showing large nanoparticle
order when $K_\mathrm{NP}\gg k_\mathrm{B}T$.  Using Eqs.~(\ref{MS_UNP}), we
obtain the approximate free energy of the nematic phase
\begin{eqnarray}
\frac{F}{N_\mathrm{LC}k_\mathrm{B}T}&=&
-\frac{J}{3k_\mathrm{B}T}S_\mathrm{LC}^2 +U_\mathrm{LC} S_\mathrm{LC}\\
&&-\ln\left[\textstyle\int_{-1}^{1}d(\cos\theta)
\exp(U_\mathrm{LC}P_2(\cos \theta))\right]\nonumber\\
&&-\frac{\nu K_\mathrm{NP}}{k_\mathrm{B}T}S_\mathrm{LC}
+\nu\ln\left(\frac{3K_\mathrm{NP}S_\mathrm{LC}}{2k_\mathrm{B}T}\right).
\nonumber
\end{eqnarray}
This free energy is equivalent to the classical Maier-Saupe free energy of
Eq.~(\ref{purefreeenergy}), except for the last two terms, which represent the
energy and entropy of well-ordered nanoparticles interacting with the liquid
crystal.  These terms are proportional to the nanoparticle concentration
$\nu=N_\mathrm{NP}/N_\mathrm{LC}$, which is small.  These terms shift the
nematic free energy, and hence shift the isotropic-nematic transition
temperature.  To find the value of the shift, we must minimize the free
energy.

To minimize the free energy, we use perturbation theory.  For this
calculation, we define the parameters
\begin{subequations}
\begin{eqnarray}
\label{MS_ULC0}U_\mathrm{LC}&=&U_\mathrm{LC}^0 +\Delta U_\mathrm{LC},\\
\label{MS_Tni0}T_\mathrm{NI}&=&T_\mathrm{NI}^0 +\Delta T_\mathrm{NI},
\end{eqnarray}
\end{subequations}
where $U_\mathrm{LC}^0$ and $T_\mathrm{NI}^0$ are the known results from the
classical Maier-Saupe free energy, given in Eqs.~(\ref{ULCpure})
and~(\ref{TNIJ}), and $\Delta U_\mathrm{LC}$ and $\Delta T_\mathrm{NI}$ are
perturbations due to addition of ferroelectric nanoparticles.  For low
nanoparticle concentrations, these perturbations should both be of order
$\nu$.  We now expand the free energy to lowest order in these pertubations,
minimize over $\Delta U_\mathrm{LC}$, and solve for $\Delta T_\mathrm{NI}$
such that the isotropic and nematic free energies are equal.  The resulting
shift in the transition temperature is
\begin{equation}
\Delta T_\mathrm{NI}=1.03\frac{\nu K_\mathrm{NP}}{k_\mathrm{B}}
=1.03\frac{\phi_\mathrm{NP}\Delta\varepsilon P^2}
{135k_\mathrm{B}\rho_\mathrm{LC}\varepsilon_0\varepsilon^2}.
\label{MS_delta_Tni0}
\end{equation}

Comparing Eqs.~(\ref{DeltaTNI}) and~(\ref{MS_delta_Tni0}), we can see that
there are two regimes for the shift in the transition temperature.  For small
interaction $K_\mathrm{NP}$ (i.e.\ the Landau regime), the shift
$\Delta T_\mathrm{NI}$ increases as $K_\mathrm{NP}^2$, but for large
$K_\mathrm{NP}$, it increases more slowly as $K_\mathrm{NP}$.  In both cases
it is proportional to the nanoparticle concentration $\nu$.  Equivalently, if
we work at fixed nanoparticle volume fraction $\phi_\mathrm{NP}$, our theory
predicts that $\Delta T_\mathrm{NI}$ will increase with the nanoparticle
material polarization $P^4$ and radius $R^3$ in the weak-interaction regime,
but it will only increase as $P^2$ and will be independent of $R$ in the
strong-interaction regime.  (It will be independent of $R$ as long as the
particles are small enough so that they do not distort the liquid-crystal
alignment.)

Our predictions for $\Delta T_\mathrm{NI}$ can be compared with the previous
predictions of Li \emph{et al.}~\cite{li06}.  They calculated that
$\Delta T_\mathrm{NI}$ should increase as the volume fraction
$\phi_\mathrm{NP}$ and as the polarization $P^2$, and should be independent of
the radius $R$.  These predictions for the scaling agree with our predictions
for the strong-interaction regime (although not for the weak-interaction
regime).  We believe that this agreement is just a coincidence, because the
theories are quite different.  One way to see the difference is through the
dependence on dielectric anisotropy $\Delta\varepsilon$:  they predict that
$\Delta T_\mathrm{NI}$ should scale as $(\Delta\varepsilon)^2$, but we
calculate that it should scale linearly with $\Delta\varepsilon$ in the
strong-interaction regime.  This difference arises because their model
considers one liquid-crystal molecule interacting through the dielectric
anisotropy $\Delta\varepsilon$ with one nanoparticle, which then interacts
through $\Delta\varepsilon$ with another liquid-crystal molecule, thus giving
an effective liquid-crystal interaction proportional to
$(\Delta\varepsilon)^2$.  By comparison, in the strong-interaction regime our
model considers the direct influence of well-ordered nanoparticles on the
liquid crystal, and hence has only one power of $\Delta\varepsilon$.

For a numerical estimate, we use typical experimental values of the parameters
$\phi_\mathrm{NP}=0.5\%$,
$P=0.26$ Cm$^{-2}$,
$R=35$ nm,
$\rho_\mathrm{LC}=2.4\times10^{27}$ m$^{-3}$,
$k_\mathrm{B}=1.38\times10^{-23}$ JK$^{-1}$,
$\epsilon_0=8.85\times10^{-12}$ C$^2$N$^{-1}$m$^{-2}$, and
$\Delta\epsilon\approx\epsilon\approx 10$.
Those parameters imply
$\nu=1\times10^{-8}$,
$K_\mathrm{NP}=1\times10^{-15}$ J,
and hence $K_\mathrm{NP}/(k_\mathrm{B} T)=2\times10^5$,
so the system is definitely in the strong-interaction regime.  Our prediction
for the shift in transition temperature is then
\begin{equation}
\Delta T_\mathrm{NI}\approx 1\text{ K}.
\label{MS_delta_Tni0_val}
\end{equation}
This value is consistent with the order of magnitude that is observed in
experiments.  Note that in this prediction we are using the bulk polarization
of the ferroelectric material BaTiO$_3$, which is $P=0.26$ Cm$^{-2}$.  In this
respect, our current estimate is different from our previous
paper~\cite{lopatina09}, where we assumed $P=0.04$ Cm$^{-2}$ because of an
understanding that the bulk polarization is reduced by surface effects in
nanoparticles.  The issue of estimating the polarization of nanoparticles is
subtle, as discussed in Ref.~\cite{shelestiuk11}.

As a final point about the phase diagram, we should mention that the model
defined by the free energy~(\ref{freeenergy}) can exhibit one additional
phase, between isotropic and nematic, which occurs if the parameter
$K_\mathrm{NP}$ is sufficiently large.  In this intermediate phase, the
nanoparticles have substantial orientational order (with $S_\mathrm{NP}$
comparable to the Maier-Saupe order parameter of 0.429), but the liquid
crystal has only very slight orientational order (with $S_\mathrm{LC}$ of
order $\nu K_\mathrm{NP}/(k_\mathrm{B} T)$).  For that reason, we might call
it a ``semi-nematic'' phase.  It is a perturbation on the pure liquid
crystal's isotropic phase, not on the nematic phase.  The semi-nematic phase
is probably an artifact of the mean-field theory used here.  It can only exist
because the very slight order of the liquid crystal mediates an aligning
interaction between the nanoparticles.  This slight orientational order is
unlikely to persist when one includes fluctuations in the liquid crystal.

\section{Kerr effect}

Apart from the phase diagram, another important issue is the response of a
liquid crystal to an applied electric field.  In the isotropic phase, an
applied field $E$ induces orientational order proportional to $E^2$, known as
the Kerr effect.  In most pure liquid crystals, the Kerr effect is quite
small, and can only be observed for very large fields.  However, in our
previous paper, we predicted that ferroelectric nanoparticles can enhance the
Kerr effect by several orders of magnitude.  We would like to assess how this
prediction is modified by the Maier-Saupe theory presented here.

In the presence of an electric field, ferroelectric nanoparticles will have
polar order along the field; i.e.\ the orientational distribution function
will no longer have a symmetry between the directions $+\mathbf{n}$ and
$-\mathbf{n}$.  Hence, we must change the nanoparticle distribution of
Eq.~(\ref{NPdistribution}) to
\begin{equation}
\varrho_\mathrm{NP}(\theta)=
\frac{e^{U^\mathrm{NP}_1 P_1(\cos\theta)+U^\mathrm{NP}_2 P_2(\cos\theta)}}
{\int_{-1}^{1}d(\cos\theta)e^{U^\mathrm{NP}_1 P_1(\cos\theta)
+U^\mathrm{NP}_2 P_2(\cos\theta)}}.
\end{equation}
Here, $U^\mathrm{NP}_1$ and $U^\mathrm{NP}_2$ are two variational parameters,
which act as effective fields on the polar and nematic order of the
nanoparticle distribution function, as described by the Legendre polynomials
$P_1(\cos\theta))$ and $P_2(\cos\theta))$, respectively.  They generate polar
and nematic order parameters, defined as
\begin{subequations}
\begin{eqnarray}
M_\mathrm{NP}&=&\frac{\int_{-1}^{1}d(\cos\theta)P_1(\cos\theta)
e^{U^\mathrm{NP}_1 P_1(\cos\theta)+U^\mathrm{NP}_2 P_2(\cos\theta)}}
{\int_{-1}^{1} d(\cos\theta)e^{U^\mathrm{NP}_1 P_1(\cos\theta)
+U^\mathrm{NP}_2 P_2(\cos\theta)}},\nonumber\\
\label{MNP}\\
S_\mathrm{NP}&=&\frac{\int_{-1}^{1}d(\cos\theta)P_2(\cos\theta)
e^{U^\mathrm{NP}_1 P_1(\cos\theta)+U^\mathrm{NP}_2 P_2(\cos\theta)}}
{\int_{-1}^{1} d(\cos\theta)e^{U^\mathrm{NP}_1 P_1(\cos\theta)
+U^\mathrm{NP}_2 P_2(\cos\theta)}}.\nonumber\\
\label{S_NP_Ee}
\end{eqnarray}
\end{subequations}
We still assume that the liquid-crystal distribution function is purely
nematic, not polar, as given by Eq.~(\ref{LCdistribution}).

The applied electric field $E$ adds two contributions to the energy of the
system,
\begin{equation}
F_\mathrm{energetic}^\mathrm{field}=-\frac{\varepsilon_0 \Delta\varepsilon}
{3\rho_\mathrm{LC}}E^2 S_\mathrm{LC}N_\mathrm{LC}
-p E M_\mathrm{NP}N_\mathrm{NP}.
\end{equation}
Here, the first term is the interaction of the field with the dielectric
anisotropy of the liquid crystal, and the second term is the interaction with
the dipole moments of the nanoparticles.  With these energetic terms, together
with the entropy of the distribution function, the free energy becomes
\begin{eqnarray}
&&\frac{F}{N_\mathrm{LC}k_\mathrm{B}T}=
-\frac{J}{3k_\mathrm{B}T}S_\mathrm{LC}^2
-\frac{\nu K_\mathrm{NP}}{k_\mathrm{B} T}S_\mathrm{LC}S_\mathrm{NP}\nonumber\\
&&\qquad-\frac{\varepsilon_0\Delta\varepsilon E^2}
{3k_\mathrm{B}T\rho_\mathrm{LC}}S_\mathrm{LC}
-\frac{\nu p E}{k_\mathrm{B}T}M_\mathrm{NP}\nonumber\\
&&\qquad+U_\mathrm{LC}S_\mathrm{LC}+\nu U^\mathrm{NP}_1 M_\mathrm{NP}
+\nu U^\mathrm{NP}_2 S_\mathrm{NP}\label{freeenergy_Ee}\\
&&\qquad-\ln\left[\textstyle\int_{-1}^{1}d(\cos\theta)
e^{U_\mathrm{LC}P_2(\cos\theta)}\right]\nonumber\\
&&\qquad-\nu\ln\left[\textstyle\int_{-1}^{1}d(\cos\theta)
e^{U^\mathrm{NP}_1 P_1(\cos\theta)+U^\mathrm{NP}_2 P_2(\cos\theta)}\right].
\nonumber
\end{eqnarray}

The next step is to minimize this free energy over all three variational
parameters $U_\mathrm{LC}$, $U^\mathrm{NP}_1$, and $U^\mathrm{NP}_2$.  For
this minimization, there are four distinct regimes of electric field, as
indicated in Fig.~2.

\begin{figure}
\includegraphics[width=3.375in,clip=true]{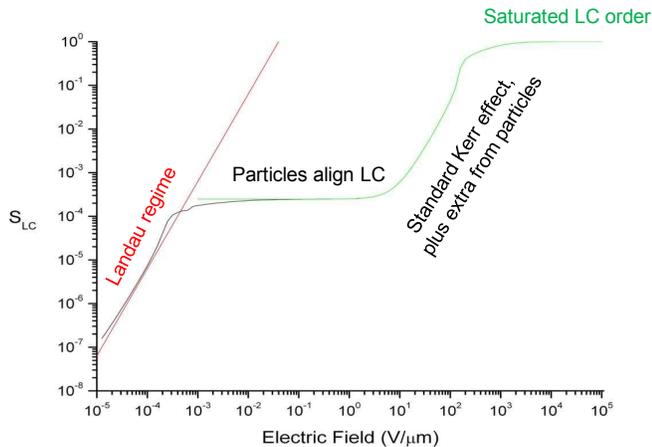}
\caption{(Color online) Four regimes of the Kerr effect, derived from a
numerical minimization of Eq.~(\ref{freeenergy_Ee}) with typical experimental
parameters.  A log-log scale is used to show all the regimes on a single
plot.}
\end{figure}

(a) If the field is sufficiently small, $E\lesssim k_\mathrm{B}T/p$, it
induces only slight order in the liquid-crystal and nanoparticle
distributions.  In that case, we can expand the free energy as a power series
in all the variational parameters.  This expansion is exactly the Landau
theory presented in our previous paper~\cite{lopatina09}.  We can then
minimize the free energy over all the variational parameters to obtain
\begin{equation}
S_\mathrm{LC}=\frac{E^2}{15 k_\mathrm{B} (T-T^*_\mathrm{doped})}
\left(\frac{\varepsilon_0\Delta\varepsilon}{\rho_\mathrm{LC}}
+\frac{\nu K_\mathrm{NP}p^2}{5(k_\mathrm{B}T)^2}\right),
\end{equation}
where $T^*_\mathrm{doped}=T^* +\Delta T_\mathrm{NI}$ is the limit of
supercooling of the nanoparticle-doped liquid crystal, combining
Eqs.~(\ref{tstar}) and~(\ref{DeltaTNI}).  In this expression, the first term
is the conventional Kerr effect without nanoparticles, and the second term is
an additional contribution due to the aligning effect of the nanoparticles.
Note that both terms are proportional to $E^2$.  With the numerical estimates
presented above, the second term is several orders of magnitude larger than
the first, and hence the nanoparticles greatly increase the Kerr effect in
this regime.

(b) For larger field, in the regime $k_\mathrm{B}T/p\lesssim E\lesssim
[\nu K_\mathrm{NP}\rho_\mathrm{LC}/(\varepsilon_0\Delta\varepsilon)]^{1/2}$,
the nanoparticle order parameters $M_\mathrm{NP}$ and $S_\mathrm{NP}$ saturate
near the maximum value of 1.  In that case, we can no longer expand the free
energy as a power series in the nanoparticle parameter, but we can still
expand it in the liquid-crystal parameter.  Minimizing the free energy then
gives
\begin{equation}
S_\mathrm{LC}=\frac{1}{15k_\mathrm{B}(T-T^*)}
\left(\frac{\varepsilon_0\Delta\varepsilon E^2}
{\rho_\mathrm{LC}}+3\nu K_\mathrm{NP}\right).
\label{Kerrregimeb}
\end{equation}
Once again, the first term is the conventional Kerr effect without
nanoparticles, and the second term is the additional contribution from the
nanoparticles, but now the second term is independent of electric field.  The
second term is still much larger than the first, and hence the Kerr effect is
approximately constant with respect to field in this regime.

(c) For even larger field,
$[\nu K_\mathrm{NP}\rho_\mathrm{LC}/(\varepsilon_0\Delta\varepsilon)]^{1/2}
\lesssim E\lesssim
[k_\mathrm{B}(T-T^*)\rho_\mathrm{LC}/(\varepsilon_0\Delta\varepsilon)]^{1/2}$,
the order parameter $S_\mathrm{LC}$ is still given by Eq.~(\ref{Kerrregimeb}),
but now the first term becomes larger than the second.  In this regime,
$S_\mathrm{LC}$ again increases as $E^2$.  It is similar to the conventional
liquid-crystal Kerr effect, but with an extra constant contribution from the
nanoparticles.

(d) For the largest field,
$[k_\mathrm{B}(T-T^*)\rho_\mathrm{LC}/(\varepsilon_0 \Delta\varepsilon)]^{1/2}
\lesssim E$, the order parameter $S_\mathrm{LC}$ saturates at the maximum
value of 1.

To get a full picture of the behavior through all these regimes, we minimize
the free energy of Eq.~(\ref{freeenergy_Ee}) numerically, using the typical
experimental parameters listed at the end of Sec.~III.  The results of this
calculation are shown by the black line in Fig.~2.  By comparison, the red
line shows the limiting case of regime~(a), and the green line shows the
approximation for regimes (b-d).  We see that the numerical solution overlaps
the limiting cases and connects them.

Note that the low-field regime~(a) is the regime where Landau theory is valid,
and it is where the nanoparticles give the greatest enhancement of the
conventional Kerr effect.  However, this regime will be difficult to observe
in experiments, because the induced order parameter $S_\mathrm{LC}$ is so
small, on the order of $10^{-4}$.  Typical optical experiments can only detect
a birefringence corresponding to $S_\mathrm{LC}$ on the order of $10^{-2}$,
which does not occur until regime~(c), which is closer to the conventional
Kerr effect.

\section{Conclusions}

In our previous paper~\cite{lopatina09}, we developed a Landau theory for the
statistical mechanics of ferroelectric nanoparticles suspended in liquid
crystals.  This theory differs from other models by considering the
orientational distribution function of the nanoparticles as well the liquid
crystal.  It shows a coupling between the nanoparticle order and the liquid
crystal order, which leads to an increase in the isotropic-nematic transition
temperature and in the Kerr effect.  In the current paper, we consider the same
physical concept, but we improve the mathematical treatment by using a
Maier-Saupe-type theory.  This theory reduces to the previous Landau theory in
the limit of weak interactions (for the isotropic-nematic transition) or weak
electric fields (for the Kerr effect).  However, it changes the results in the
opposite limit, when the order parameters begin to saturate.  For that reason,
the new theory should make more accurate predictions for experiments.

In general, the concept of coupled orientational distribution functions should
be useful for many other systems beside ferroelectric nanoparticles in liquid
crystals.  For example, it applies to any type of nonspherical colloidal
particles, such as carbon nanotubes, in a liquid-crystal solvent.  It also
applies to two distinct species of nonspherical colloids suspended in an
isotropic solvent, which could have a coupled ordering transition.  Such
systems would provide further opportunities to investigate the theory
presented here.

\acknowledgments

We would like to thank Y. Reznikov and J. L. West for many helpful
discussions.  This work was supported by NSF Grant DMR-0605889.

\end{document}